\documentclass{PoS}

\usepackage{amsfonts}
\usepackage{amsmath}
\usepackage{amssymb}
\usepackage{wrapfig}

\newcommand{\beq}{\begin{equation}}
\newcommand{\eeq}{\end{equation}}
\newcommand{\be}{\begin{equation}}
\newcommand{\ee}{\end{equation}}

\newcommand{\G}{\mathrm{SL}(2) \times \mathbb{R}^+}
\newcommand{\Edd}{E_{d(d)}}

\newcommand{\dd}{\mathrm{d}}		
			
\newcommand{\cH}{\mathcal{H}}			
\newcommand{\MM}{\mathcal{M}}
\newcommand{\gM}{\mathcal{M}}
\newcommand{\LL}{\mathcal{L}}			

\newcommand{\hR}{\hat{R}}
\newcommand{\bz}{\bar{z}}

\newcommand{\hx}{\hat{x}}
\newcommand{\hmu}{{\hat{\mu}}}
\newcommand{\hnu}{{\hat{\nu}}}

\newcommand{\hrho}{\hat{\rho}}
\newcommand{\hlambda}{\hat{\lambda}}

\newcommand{\gL}{\LL}

\newcommand{\1}{{\mu_1}}

\newcommand{\3}{{\mu_3}}
\newcommand{\4}{{\mu_4}}

\newcommand{\6}{{\mu_6}}
\newcommand{\7}{{\mu_7}}

\newcommand{\mt}{{\mu_{10}}}

\newcommand{\Aa}{A}
\newcommand{\Ab}{B}
\newcommand{\Ac}{C}
\newcommand{\Ad}{D}
\newcommand{\Ae}{E}
\newcommand{\Af}{F}

\newcommand{\Fa}{\mathcal{F}}
\newcommand{\Fb}{\mathcal{H}}
\newcommand{\Fc}{\mathcal{J}}
\newcommand{\Fd}{\mathcal{K}}
\newcommand{\Fe}{\mathcal{L}}

\title{$\G$ Exceptional Field Theory: \\ An Action for F-Theory}

\ShortTitle{$\G$ Exceptional Field Theory}

\author{\speaker{Felix J. Rudolph}\\
       Arnold-Sommerfeld-Center, LMU, Munich\\
       E-mail: \email{felix.rudolph@lmu.de}}

\abstract{
Exceptional Field Theory employs an extended spacetime to make supergravity fully covariant under the U-duality groups of M-theory. The 12-dimensional EFT associated to the group $\G$ together with its action is presented. Demanding the closure of the algebra of local symmetries leads to a constraint, known as the section condition, that must be imposed on all fields. This constraint has two inequivalent solutions, one giving rise to 11-dimensional supergravity and the other leading to Type IIB supergravity and F-theory. Thus $\G$ Exceptional Field Theory contains both F-theory and M-theory in a single 12-dimensional formalism.}

\FullConference{Corfu Summer Institute 2016 "School and Workshops on Elementary Particle Physics and Gravity"\\
         31 August - 23 September, 2016\\
         Corfu, Greece}
         
\begin{document}

\section{Exceptional Field Theory}

The low energy effective description of M-theory is known to be 11-dimensional supergravity with the coupling of the type IIA string promoted to the radius of the eleventh dimension. The natural extension of this idea to the type IIB string gives rise to F-theory \cite{Vafa:1996xn} where the complex coupling in the IIB theory is taken to have its origin in the complex modulus of a torus fibred over the usual ten dimensions of the type IIB string theory. Thus by definition, F-theory is the 12-dimensional lift of type IIB string theory. 

The status of this 12-dimensional theory has been somewhat different to that of its IIA spouse with no direct 12-dimensional description in terms of an action and fields that reduce to the IIB theory. Indeed, there is no 12-dimensional supergravity and thus no limit in which the eleventh and twelfth dimension can be taken to be ``large'', unlike in the M-theory case. The emphasis has thus largely been on using algebraic geometry to describe F-theory compactifications such that now F-theory is synonymous with the study of elliptically fibred Calabi-Yau manifolds. 

The complex coupling of the IIB theory is naturally acted on by an $\mathrm{SL}(2)$ S-duality, which is a symmetry of the theory. F-theory thus provides a geometric interpretation of this duality. The idea of reimagining a duality in a geometric way has reappeared in the form of double field theory (DFT) \cite{Siegel:1993,Hull:2009mi} and exceptional field theory (EFT) \cite{Hohm:2013pua}. Perhaps the key development that allows for the construction of these theories is that the new geometry is not a conventional one, but an ``extended geometry'' based on the idea of ``generalised geometry'' \cite{Hitchin:2004ut, Gualtieri:2003dx}. For instance, a key role is played by a ``generalised metric'', in place of the torus modulus, and one introduces an extended space with a novel ``generalised diffeomorphism'' symmetry. 

The primary idea behind exceptional field theory is to make the exceptional symmetries of eleven-dimensional supergravity manifest. The appearance of the exceptional groups in dimensionally reduced supergravity theories was first discussed in \cite{Cremmer:1998}. In EFT one first performs a decomposition of eleven-dimensional supergravity -- but with no reduction or truncation -- into an $(11-d) \times d$ split. Then one supplements the $d$ so-called ``internal'' directions with additional coordinates to linearly realize the exceptional symmetries. That is one extends the eleven dimensions of supergravity to
\begin{equation}
M^{11} = M^{11-d} \times M^d \longrightarrow M^{11-d} \times M^{\dim \Edd}
\end{equation}
where $\dim \Edd$ is the dimension of the relevant representation of the exceptional group $\Edd$ and $M^{\dim \Edd}$ is a coset manifold that comes equipped with the coset metric of $\Edd/H$ (where $H$ is the maximally compact subgroup of $\Edd$). 

The U-duality groups are related to the embedding of the eleven dimensions in the extended space. The combination of $p$-form gauge transformations and diffeomorphism give rise to a continuous local $\Edd$ symmetry. This however is not U-duality which is a global discrete symmetry that only occurs in the presence of isometries. (See \cite{Berman:2014jba} for the equivalent discussion of T-duality in DFT). 

Crucially however there is also a \emph{physical section condition} that provides a constraint in EFT that restricts the coordinate dependence of the fields to a subset of the dimensions and thus there naturally appears a physical submanifold which is identified as the usual spacetime. When there are no isometries present this section condition constraint produces a canonical choice of how spacetime is embedded in the extended space. However, in the presence of isometries there is an ambiguity in how one identifies the submanifold in the extended space. This ambiguity is essentially the origin of U-duality with different choices of spacetime associated to U-duality related descriptions. (This is discussed in detail for the case of DFT in \cite{Berman:2014jsa} and for EFT in \cite{Berman:2014hna}).

In EFT a solution to the $\Edd$ section condition will provide either a $d$-dimensional space or a $\left(d-1\right)$-dimensional space (where crucially the $d-1$ solution is not a subspace of the $d$-dimensional space). The two solutions are distinct (and not related by any element of $\Edd$). The $d$-dimensional solution is associated to the M-theory description and the $\left(d-1\right)$-dimensional solution is associated to the type IIB description. 

A completely generic solution that solves the section condition will be in one set or the other and one will be able to label it as an M or IIB solution. However, if there are {\bf{two}} isometries in the M-theory solution then again we have an ambiguity and one will be able interpret the solution in terms of IIB section with {\bf{one}} isometry. This ambiguity gives the F-theory/M-theory duality. It is the origin of how M-theory on a torus is equivalent to IIB on a circle \cite{Schwarz:1995jq}. Thus in summary the F-theory/M-theory duality is an ambiguity in the identification of spacetime that occurs when there are two isometries in an M-theory solution.

The relation between M-theory/Type IIA on the one side and F-theory/Type IIB on the other is clearest in the ``smallest'' EFT where $d=2$, the U-duality group is $\G$ and just one extra dimension is needed for the construction. This 12-dimensional theory has been constructed in \cite{Berman:2015rcc} where it has been established that the $\G$ EFT provides a local action for F-theory (also see \cite{Linch:2015fca}). These results where presented at the ``Workshop on Geometry and Physics" at Ringberg Castle (20 - 25 November 2016) and will be summarised here.

\section{The $\G$ EFT}
The goal is to provide an overview of the exceptional field theory in $9+3$ dimensions and to describe the general features and the setup of the extended space. Then the field content and the action are presented. The details of the construction can be found in the original paper \cite{Berman:2015rcc}. The theory we will discuss may be thought of as a 12-dimensional theory with a $9+3$ split of the coordinates, so that we have
\begin{itemize}
\item nine ``external'' coordinates, $x^\mu$,
\item three ``extended'' coordinates, $Y^M$ that live in the $\mathbf{2}_{1} \oplus \mathbf{1}_{-1}$ reducible representation of $\G$ (where the subscripts denote the weights under the $\mathbb{R}^+$ factor). To reflect the reducibility of the representation\footnote{The reducibility of the coordinate representation is not a feature of higher rank duality groups.} we further decompose the coordinates $Y^M = ( y^\alpha, y^s)$ where $\alpha=1,2$ transforms in the fundamental of $\mathrm{SL}(2)$, and $s$ stands for ``singlet''.
\end{itemize}
The fields and symmetry transformation parameters of the theory can in principle depend on all of these coordinates. However, as always happens in exceptional field theory and double field theory, there is a consistency condition which reduces the dependence on the extended coordinates. This condition is usually implemented as the section condition, which directly imposes that that the fields cannot depend on all extended coordinates. In our case, it takes the form
\be
\partial_\alpha \partial_s = 0 \,,
\label{eq:constraint}
\ee
with the derivatives to be thought of as acting on any field or pair of fields, so that we require both $\partial_\alpha \partial_s \mathcal{O} =0$ and $\partial_\alpha \mathcal{O}_1 \partial_s \mathcal{O}_2 + \partial_\alpha \mathcal{O}_2 \partial_s \mathcal{O}_1 = 0$. The origin of the section condition is the requirement that the algebra of symmetries closes and is discussed in more detail in \cite{Berman:2015rcc}.%\footnote{In fact, one can alternatively use a generalised Scherk-Schwarz ansatz to obtain a set of weaker conditions that replace the above section condition with a set of constraints on twist matrices which in principle depend on all the extended coordinates \cite{Aldazabal:2011nj,Geissbuhler:2011mx,Grana:2012rr,Berman:2012uy}. This would be an interesting possibility to apply here, but we do not pursue this in the present paper.}

The action we will present has a manifest invariance under a global $\G$ symmetry, acting on the indices $M=(\alpha,s)$ in an obvious way. In addition, the exceptional field theory is invariant under a set of local symmetries.

\subsection{Local and global symmetries} 

Alongside the introduction of the extended coordinates $Y^M$ one constructs so called ``generalised diffeomorphisms". In the higher rank groups, these give a unified description of ordinary diffeomorphisms together with the $p$-form gauge transformations. Although the group $\G$ is too small for the $p$-form gauge transformations to play a role here, the generalised diffeomorphisms provide a combined description of part of the ordinary local symmetries of IIB and 11-dimensional supergravity.

The generalised diffeomorphisms, generated by a generalised vector $\Lambda^M$, act as a local $\G$ action, called the generalised Lie derivative $\gL_\Lambda$. These act on a vector, $V^M$ of weight $\lambda_V$ in a form which looks like the usual Lie derivative $L_\Lambda$ plus a modification involving the so-called ``Y-tensor'' which in the case of $\G$ \cite{Wang:2015hca} is symmetric on both upper and lower indices and has the only non-vanishing components $Y^{\alpha s}{}_{\beta s} = \delta^\alpha_\beta$
\begin{equation}
\delta_\Lambda V^M \equiv \gL_\Lambda V^M
	= \Lambda^N\partial_NV^M - V^N\partial_N\Lambda^M + Y^{MN}{}_{KL}\partial_N\Lambda^K V^L
		+ (\lambda_V+\omega) \partial_N \Lambda^N V^M \, .
\end{equation}
There is also a universal weight term, $+ \omega \partial_N \Lambda^N V^M$. The constant $\omega$ depends on the number $n=11-d$ of external dimensions as $\omega = - \frac{1}{n-2}$ and for us $\omega = - 1/7$. The gauge parameters themselves are chosen to have specific weight $\lambda_\Lambda = 1/7$, which cancels that arising from the $\omega$ term. 

The transformation rules for the components $V^\alpha$ and $V^s$ of a generalised vector $V^M$ are 
\begin{equation}
 \begin{split}
  \gL_\Lambda V^\alpha &= \Lambda^M \partial_M V^\alpha - V^\beta \partial_\beta \Lambda^\alpha - \frac17 V^\alpha \partial_\beta \Lambda^\beta + \frac67 V^\alpha \partial_s \Lambda^s \,, \\
  \gL_\Lambda V^s &= \Lambda^M \partial_M V^s + \frac67 V^s \partial_\beta \Lambda^\beta - \frac87 V^s \partial_s \Lambda^s \,.
 \end{split}
\label{eq:gldcpts} 
\end{equation}
Then by requiring the Leibniz property for the generalised Lie derivative, we can derive the transformation rules for tensors in other representations of $\G$, such as the generalised metric $\gM_{MN}$. (The form fields must be treated separately, see \cite{Berman:2015rcc}.)

%
%\be
%\delta_\Lambda V^M \equiv
%\mathcal{L}_\Lambda V^M 
%= \Lambda^N \partial_N V^M 
%- V^N \partial_N \Lambda^M 
%+ Y^{MN}{}_{PQ} \partial_N \Lambda^P V^Q 
%+ ( \lambda_V + \omega) \partial_N \Lambda^N V^M \,.
%\label{eq:gld}
%\ee
%This modification is universal for all exceptional field theory generalised Lie derivatives \cite{Berman:2012vc} and is built from the invariant tensors of the duality group. In the case 
%\be
%Y^{\alpha s}{}_{\beta s} = \delta^\alpha_\beta \, .
%\label{eq:Y}
%\ee
%There is also a universal weight term, $+ \omega \partial_N \Lambda^N V^M$. The constant $\omega$ depends on the number $n=11-d$ of external dimensions as $\omega = - \frac{1}{n-2}$ and for us $\omega = - 1/7$. The gauge parameters themselves are chosen to have specific weight $\lambda_\Lambda = 1/7$, which cancels that arising from the $\omega$ term. 

In conventional geometry, diffeomorphisms are generated by the Lie derivative and form a closed algebra under the Lie bracket. The algebra of generalised diffeomorphisms involves the $E$-bracket,
\be
[ U,V]_E = \frac{1}{2} \left( \mathcal{L}_U V - \mathcal{L}_V U \right) \,.
\ee
The condition for closure of the algebra is
\be
\mathcal{L}_U \mathcal{L}_V - 
\mathcal{L}_V \mathcal{L}_U 
= \mathcal{L}_{[U,V]_E}  
\ee
which does not happen automatically. A universal feature in all exceptional field theories is that we need to impose the section condition \cite{Berman:2012vc} so the algebra closes. This is the constraint \eqref{eq:constraint} given above.

%In the above we have only treated the infinitesimal, local $\G$ symmetry. This should be related to finite $\G$ transformations by exponentiation. The relation between the exponentiated generalised Lie derivative and the finite transformations are quite nontrivial due to the presence of the section condition. For double field theory there are now are series of works dealing with this issue \cite{Hohm:2012gk,Park:2013mpa, Berman:2014jba,Hull:2014mxa,Naseer:2015tia,Rey:2015mba} and recently the EFT case has been studied in \cite{Chaemjumrus:2015vap}.

The other diffeomorphism symmetry of the action consists of external diffeomorphisms, para-metrised by vectors $\xi^\mu$. These are given by the usual Lie derivative
\be
\delta_\xi V^\mu \equiv L_\xi V^\mu 
= \xi^\nu D_\nu V^\mu - V^\nu D_\nu \xi^\mu 
+ \hat\lambda_V D_\nu \xi^\nu V^\mu \,,
\ee
with partial derivatives replaced by the derivative $D_\mu$ which is covariant under internal diffeomorphisms, and explicitly defined by
\be
D_\mu = \partial_\mu - \delta_{\Aa_\mu} \,.
\label{eq:Dmu} 
\ee
where $\Aa_\mu$ is the first of the gauge fields introduced in the tensor hierarchy of the next subsection. The weight $\hat\lambda_V$ of a vector under external diffeomorphisms is independent of the weight $\lambda_V$ under generalised diffeomorphisms.

For this to work, the gauge vector $\Aa_\mu$ must transform under generalised diffeomorphisms as
\be
\delta_\Lambda \Aa_\mu{}^M= D_\mu \Lambda^M\,.
\label{eq:deltaLambdaAa} 
\ee
The external metric and form fields then transform under external diffeomorphisms in the usual manner given by the Leibniz rule, while the generalised metric is taken to be a scalar, $\delta_\xi \gM_{MN} = \xi^\mu D_\mu \gM_{MN}$.

\subsection{Field Content}

The field content of the theory is as follows. The metric-like degrees of freedom are
\begin{itemize}
\item an ``external'' metric, $g_{\mu \nu}$, 
\item a generalised metric, $\gM_{MN}$ which parametrises the coset $(\G)/\mathrm{SO}(2)$. (From the perspective of the ``external'' nine dimensions, this metric will correspond to the scalar degrees of freedom.) The reducibility of the $Y^M$ coordinates implies that the generalised metric is reducible and thus may be decomposed as, \be \gM_{MN}=\gM_{\alpha \beta} \oplus \gM_{ss} \, . \ee
\end{itemize}
The coset $(\G)/\mathrm{SO}(2)$ implies we have just three degrees of freedom described by the generalised metric. This means that $\gM_{ss}$ must be related to $\det \gM_{\alpha \beta}$. One can thus define $\gM_{\alpha \beta}$ such that
\be
\cH_{\alpha \beta} \equiv (\gM_{ss})^{3/4} \gM_{\alpha \beta} 
\label{eq:unitdetgenmetric}
\ee
has unit determinant. The rescaled metric $\cH_{\alpha \beta}$ and $\gM_{ss}$ can then be used as the independent degrees of freedom when constructing the theory. This unit determinant matrix $\cH_{\alpha \beta}$ will appear naturally in the Type IIB/F-theory description. 

In addition, we have a hierarchy of gauge fields, similar to the tensor hierarchy of gauged supergravities \cite{deWit:2005hv}. These are form fields with respect to the external directions and transform in different representations of the duality group:
\be
\begin{array}{|c|ll|ll|} \hline
\mathrm{Representation} & & \mathrm{Gauge\ potential} & & \mathrm{Field}\,\, \mathrm{strength} \\ \hline
\mathbf{2}_1 \oplus \mathbf{1}_{-1} & & \Aa_\mu{}^M & & \Fa_{\mu \nu}{}^M \\
\mathbf{2}_0 & & \Ab_{\mu \nu}{}^{\alpha s} & & \Fb_{\mu \nu \rho}{}^{\alpha s} \\ 
\mathbf{1}_1 & & \Ac_{\mu \nu \rho}{}^{[\alpha \beta] s} & & \Fc_{\mu \nu \rho \sigma}{}^{[\alpha \beta] s} \\
\mathbf{1}_0 & & \Ad_{\mu \nu \rho \sigma}{}^{[\alpha \beta] ss } & & \Fd_{\mu \nu \rho \sigma \lambda}{}^{[\alpha \beta]ss} \\
\mathbf{2}_1 & & \Ae_{\mu \nu \rho \sigma \kappa}{}^{\gamma [\alpha \beta] ss } & & \Fe_{\mu \nu \rho \sigma \kappa\lambda}{}^{\gamma [\alpha \beta]ss} \\
\mathbf{2}_0 \oplus \mathbf{1}_2 & & \Af_{\mu \nu \rho \sigma \kappa \lambda}{}^{M} & & \mathrm{not}\,\,\mathrm{needed} \\  \hline
\end{array}
\label{eq:forms}
\ee

These gauge fields also transform under the generalised diffeomorphisms and external diffeomorphisms described above, as well as various gauge symmetries of the tensor hierarchy (see \cite{Berman:2015rcc} for details). The field strengths are defined such that the fields transform covariantly under generalised diffeomorphisms, i.e. according to their index structure and the rules given above, and are gauge invariant under a hierarchy of interrelated gauge transformations as detailed in \cite{Berman:2015rcc}. 

We also need to specify the weight $\lambda$ of each object. It is conventional to choose the generalised metric to have weight zero under generalised diffeomorphisms. Meanwhile, the sequence of form fields $\Aa, \Ab, \Ac,\dots$ are chosen to have weights $\lambda_\Aa = 1/7$, $\lambda_{\Ab} = 2/7$, $\lambda_{\Ac} = 3/7$ and so on. Finally, we take the external metric $g_{\mu \nu}$ to be a scalar of weight $2/7$.

%The expressions for the field strengths are schematically
%\be
%\begin{split} 
%\mathcal{F}_{\mu \nu}{} & = 2 \partial_{[\mu} A_{\nu]} +  \dots + \hd\Ab_{\1\2}\, \\
%\Fb_{\1\2\3} &= 3\D_{[\1}\Ab_{\2\3]} +  \dots + \hd\Ac_{\1\2\3} \,, \\
%\Fc_{\1\ldots\1\4} &= 4\D_{[\1}\Ac_{\2\3\4]}  + \dots + \hd\Ad_{\1\ldots\4} \,, \\
%\Fd_{\1\ldots\5} &= 5\D_{[\1}\Ad_{\2\ldots\5]} + \dots + \hd\Ae_{\1\ldots\5} \,, \\
%\Fe_{\1\ldots\6} &= 6\D_{[\1}\Ae_{\2\ldots\6]}  + \dots + \hd\Af_{\1\ldots\6} \,,
%\end{split} 
%\ee
%where for the $p$-form field strengths the terms indicated by dots involve quadratic and higher order of field strengths. We also see that there is always a linear term, shown, of the gauge field at next order, under a particular nilpotent derivative $\hd$ defined in \cite{Berman:2015rcc}. The derivative, $D_\mu$ that appears is a covariant derivative for the generalised diffeomorphisms, as described below. The detailed definitions of the field strengths are also in \cite{Berman:2015rcc}. 

\subsection{The Action}

The presence of the two kinds of diffeomorphism symmetries may be used to fix the action up to total derivatives. The resulting general form of the action, which is common to all exceptional field theories is given schematically as follows, 
\be
S = \int \dd^{9} x \dd^3 Y \sqrt{g} \left( \hR + 
\LL_{skin} 
+ \LL_{gkin} 
+ \frac{1}{\sqrt{g}} \LL_{top} 
+ V 
\right) \,.
\label{eq:S} 
\ee
The constituent parts are (omitting total derivatives):
\begin{itemize}
\item the ``covariantised'' external Ricci scalar, $\hR$, which is 
\be
\hR = 
\frac{1}{4} g^{\mu \nu} D_\mu g_{\rho \sigma} D_\nu g^{\rho \sigma} 
- \frac{1}{2} g^{\mu \nu} D_\mu g^{\rho \sigma} D_\rho g_{\nu \sigma} 
+ \frac{1}{4} g^{\mu \nu} D_\mu \ln g D_\nu \ln g + \frac{1}{2}  D_\mu \ln g D_\nu g^{\mu \nu} \,.
\ee 
\item a kinetic term for the generalised metric containing the scalar degrees of freedom
\be
\mathcal{L}_{skin}
= -\frac{7}{32} g^{\mu \nu} D_\mu \ln \gM_{ss} D_\nu \ln \gM_{ss} 
+ \frac{1}{4}  g^{\mu \nu} D_\mu \cH_{\alpha \beta} D_\nu \cH^{\alpha \beta} \, ,
\ee

\item kinetic terms for the gauge fields
\be
\begin{split}
\mathcal{L}_{gkin} & = 
- \frac{1}{2\cdot 2!} \gM_{MN} \mathcal{F}_{\mu \nu}{}^M \mathcal{F}^{\mu \nu N} 
 - \frac{1}{2\cdot 3!} \gM_{\alpha \beta} \gM_{ss} \mathcal{H}_{\mu \nu \rho}{}^{\alpha s} \mathcal{H}^{\mu \nu \rho \beta s} 
\\ & \qquad - \frac{1}{2\cdot 2! 4!} \gM_{ss} \gM_{\alpha \gamma} \gM_{\beta \delta} \mathcal{J}_{\mu \nu \rho \sigma}{}^{[\alpha \beta] s} \mathcal{J}^{\mu \nu \rho \sigma [\gamma \delta]s} 
\,.
\end{split}
\ee
We do not include kinetic terms for all the form fields appearing in \eqref{eq:forms}. As a result, not all the forms are dynamical. We will discuss the consequences of this below. 
\item a topological or Chern-Simons like term which is not manifestly gauge invariant in 9+3 dimensions. In a standard manner however we may write this term in a manifestly gauge invariant manner in 10+3 dimensions as
\begin{equation}
\begin{aligned}
  S_{top} &= \kappa \int d^{10}x\, d^3Y\, \varepsilon^{\1\ldots\mt}  \frac{1}{4}  \epsilon_{\alpha \beta} \epsilon_{\gamma \delta}  \left[ 
  \frac{1}{5}  \partial_s \Fd_{\mu_1 \dots \mu_5}{}^{\alpha \beta ss} \Fd_{\mu_6 \dots \mu_{10}}{}^{\gamma \delta ss} \right. \\ 
  & \qquad \left. - \frac{5}{2} \Fa_{\mu_1 \mu_2}{}^s \Fc_{\3\ldots\6}{}^{\alpha \beta s} \Fc_{\7\ldots\mt}{}^{\gamma \delta}  \right. \\
    & \qquad \left. + \frac{10}{3}2 \Fb_{\mu_1 \dots \mu_3}{}^{\alpha s}\Fb_{\4\ldots\6}{}^{\beta s} \Fc_{\7 \ldots \mt}{}^{\gamma \delta}  \right] \,.
\end{aligned} 
\label{eq:ToptTerm}
\end{equation}
The index $\mu$ is treated to an abuse of notation where it is simultaneously 10- and 9-dimen-sional. (This extra dimension is purely a notational convenience and is unrelated to the extra coordinates present in $Y^M$.) The above term is such that its variation is a total derivative and so can be written again in the correct number of dimensions. The overall coefficient $\kappa$ is found to be $\kappa = +\frac{1}{48\cdot 5!}$. % PLUS
\item a scalar potential
\be
\begin{split}
V & = 
\frac{1}{4} \gM^{ss}\left( 
 \partial_s \cH^{\alpha \beta} \partial_s \cH_{\alpha \beta} 
 + \partial_s g^{\mu \nu} \partial_s g_{\mu \nu} + \partial_s \ln g \partial_s \ln g 
 \right) 
 \\ & 
 +  \frac{9}{32}  \gM^{ss} \partial_s \ln \gM_{ss} \partial_s \ln \gM_{ss}
- \frac{1}{2}  \gM^{ss} \partial_s \ln \gM_{ss} \partial_s \ln g 
\\ &  +
\gM_{ss}^{3/4} \Bigg[ 
\frac{1}{4} \cH^{\alpha \beta} \partial_\alpha \cH^{\gamma \delta} \partial_\beta \cH_{\gamma \delta} 
+ \frac{1}{2} \cH^{\alpha \beta} \partial_\alpha \cH^{\gamma \delta} \partial_\gamma \cH_{\delta \beta} 
+ \partial_\alpha \cH^{\alpha \beta} \partial_\beta \ln \left(  g^{1/2} \gM_{ss}^{3/4}  \right)  
\\ & 
+ \frac{1}{4} \cH^{\alpha \beta} \left( \partial_\alpha g^{\mu \nu} \partial_\beta g_{\mu \nu} 
+   \partial_\alpha \ln g \partial_\beta \ln g
+ \frac{1}{4} \partial_\alpha \ln \gM_{ss} \partial_\beta \ln \gM_{ss}  + \frac{1}{2} \partial_\alpha \ln g \partial_\beta \ln \gM_{ss} 
 \right) 
\Bigg] \,.
\end{split}
\ee
\end{itemize}

This theory expresses the dynamics of 11-dimensional supergravity and 10-dimensional type IIB supergravity in a duality covariant way. In order to do so, we have actually increased the numbers of degrees of freedom by simultaneously treating fields and their electromagnetic duals on the same footing. This can be seen in the collection of form fields in \eqref{eq:forms}. For instance, although 11-dimensional supergravity contains only a three-form, here we have additional higher rank forms which can be thought of as corresponding to the six-form field dual to the three-form. 

The action for the theory deals with this by not including kinetic terms for all the gauge fields. The field strength $\Fd_{\mu\nu\rho\sigma\kappa}$ of the gauge field $D_{\mu \nu \rho \sigma}$ only appears in the topological term \eqref{eq:ToptTerm}. The field $D_{\mu\nu\rho\sigma}$ in fact also appears in the definition of the field strength $\Fc_{\mu \nu \rho}$, under a $\partial_M$ derivative. One can show that the equation of motion for this field is 
\be
\partial_s \left( \frac{\kappa}{2} \epsilon^{\mu_1 \dots \mu_9} \epsilon_{\alpha \beta} 
\epsilon_{\gamma \delta} \Fd_{\mu_5 \dots \mu_9}{}^{\gamma \delta ss} 
-  e \frac{1}{48} \gM_{ss}  \gM_{\alpha \gamma} \gM_{\beta \delta} \Fc^{\mu_1 \dots \mu_4 \gamma \delta s} \right) = 0  \,.
\label{eq:Deom} 
\ee
The expression in the brackets should be imposed as a duality relation relating the field strength $\Fd_{\mu \nu \rho \sigma \lambda}$ to $\Fc_{\mu \nu \rho \sigma}$, and hence removing seemingly extra degrees of freedom carried in the gauge fields which are actually just the dualisations of physical degrees of freedom. The above relation is quite important -- for instance the proof that the EFT action is invariant under diffeomorphisms is only obeyed if it is satisfied. 

As for the remaining two gauge fields, the equation of motion following from varying with respect to $E_{\mu \nu \rho \sigma \kappa}$ is trivially satisfied (it only appears in the field strength $\Fd_{\mu \nu \rho \sigma \kappa}$), while $F_{\mu \nu \rho \sigma \kappa \lambda}$ is entirely absent from the action.

Each individual term in the general form of the action \eqref{eq:S} is separately invariant under generalised diffeomorphisms and gauge transformations. The external diffeomorphisms though mix the various terms and so by requiring invariance under these transformations one may then fix the coefficients of the action.

%An alternative derivation of the generalised Lie derivative is the following. 
%We consider a general ansatz for the generalised Lie derivative acting on elements in the 2-dimensional and singlet representation
%\begin{equation}
% \begin{split}
%  \gL_\Lambda V^\alpha &= \Lambda^M \partial_M V^\alpha - V^\beta \partial_\beta \Lambda^\alpha + a V^\alpha \partial_\beta \Lambda^\beta + b V^\alpha \partial_\beta \Lambda^\beta \,, \\
%  \gL_\Lambda V^s &= \Lambda^M \partial_M V^s + c V^s \partial_\beta \Lambda^\beta + d V^s \partial_s \Lambda^s \,.
% \end{split}
%\end{equation}
%We can fix the coefficients $a$, $b$, $c$, $d$ as follows. We require a singlet $\Delta_s$ and $\epsilon_{\alpha\beta}$ to define an invariant, i.e.
%\begin{equation}
% \gL_\Lambda \left(\epsilon_{\alpha\beta} \Delta_s\right) = 0 \,.
%\end{equation}
%This property allows us to define the unit-determinant generalised metric \eqref{eq:unitdetgenmetric}. Furthermore, we require that the algebra of generalised Lie derivatives closes subject to a section condition. Requiring this to allow for two inequivalent solutions then fixes the coefficients $a$, $b$, $c$, $d$ above and -- up to a redefinition -- reproduces \eqref{eq:gldcpts}. This definition of the generalised Lie derivative fits in the usual pattern of generalised diffeomorphism in EFT described by the Y-tensor \cite{Berman:2012vc} deformation of the Lie derivative. 

\section{Relation to M-Theory and F-Theory}

The $\G$ exceptional field theory is equivalent to 11-dimensional and 10-dimensional IIB supergravity, in a particular splitting inspired by Kaluza-Klein reductions. The details of this split and the precise relationships between the fields of the exceptional field theory $(g_{\mu\nu}, \MM_{MN},$ $A_\mu{}^M,B_{\mu\nu}{}^{\alpha,s},C_{\mu\nu\rho}{}^{\alpha\beta,s},D_{\mu\nu\rho\sigma}{}^{\alpha\beta, ss})$ on the one side and those of M-theory $(G_{\mu\nu},\gamma_{\alpha\beta}, A_\mu{}^\alpha, \hat{C}_{\hmu\hnu\hrho})$ and type IIB $(G_{\mu\nu},\phi, A_\mu{}^s, \varphi, C_0, \hat{C}_{\hmu\hnu}{}^\alpha, \hat{C}_{\hmu\hnu\hrho\hlambda})$ on the other side are given in \cite{Berman:2015rcc}. In summary, these dictionaries can be written as follows (note that $\hx^\hmu$ is either $(x^\mu,y^\alpha)$ or $(x^\mu,y^s)$ in the two cases).

For M-theory, we impose the M-theory section condition, $\partial_s = 0$. Thus, the fields of our theory depend on the coordinates $x^\mu$ and $y^\alpha$, which are taken to be the coordinates of 11-dimensional supergravity in a $9+2$ splitting. Then one has for the degrees of freedom coming from the spacetime metric, with the Kaluza-Klein vector $A_\mu{}^\alpha = \gamma^{\alpha \beta} G_{\mu \beta}$, and the gauge fields ($\gamma=\det\gamma_{\alpha\beta}$)
\begin{align}
	g_{\mu\nu} &= \gamma^{1/7} \left( G_{\mu \nu} 
			- \gamma_{\alpha \beta} A_\mu{}^\alpha A_\nu{}^\beta \right) \,, &
	A_\mu{}^s &= \frac12 \epsilon^{\alpha\beta} \hat{C}_{\mu\alpha\beta} \,, \notag \\		
	\cH_{\alpha\beta} &= \gamma^{-1/2} \gamma_{\alpha\beta} \,, &
	B_{\mu\nu}{}^{\alpha,s} &= \epsilon^{\alpha\beta} \hat{C}_{\mu\nu\beta} 
			+ \frac12 \epsilon^{\beta\gamma} A_{[\mu}{}^\alpha \hat{C}_{\nu]\beta\gamma} \,, \label{eq:MEFT}\\
	{\cal M}_{ss} &= \gamma^{-6/7} \,, &	
  	C_{\mu\nu\rho}{}^{\alpha\beta,s} &= \epsilon^{\alpha\beta} \left( \hat{C}_{\mu\nu\rho} 
  			- 3 A_{[\mu}{}^{\gamma} \hat{C}_{\nu\rho]\gamma} 
  			+ 2 A_{[\mu}{}^{\gamma} A_{\nu]}{}^{\delta} \hat{C}_{\rho\gamma\delta} \right) \,. \notag
\end{align}

For type IIB, we impose the IIB section, $\partial_\alpha = 0$. The fields then depend on the coordinates $x^\mu$ and $y^s$, which become the coordinates of 10-dimensional type IIB supergravity in a $9+1$ split. The degrees of freedom from the spacetime metric with $\phi \equiv G_{ss}$ and the Kaluza-Klein vector $A_\mu{}^s = \phi^{-1} G_{\mu s}$ parametrise the external metric and the components of the generalised metric as
\begin{align}
g_{\mu\nu} &= \phi^{1/7} \left( G_{\mu\nu} - \phi A_\mu{}^s A_\nu{}^s \right) &
\MM_{\alpha\beta} &= \phi^{-6/7} \cH_{\alpha\beta} \, ,  &
\MM_{ss} &= \phi^{8/7} \, .
\label{eq:genmetricIIB}
\end{align}
The Kaluza-Klein vector $A_\mu{}^s$ can be identified as the $s$ component of the gauge field $A_\mu{}^M$. %One can then verify the reduction of the scalar potential, scalar kinetic terms and external Ricci scalar, and verify that they give the expected reduction of the Einstein-Hilbert term, \eqref{eq:EHresult} and scalar kinetic terms \eqref{eq:Skinred}. 
The parametrisation of $\cH_{\alpha \beta}$ in terms of the axio-dilaton $\tau = C_0 + i e^{-\varphi}$ is given by 
\begin{equation}
\cH_{\alpha\beta} = \frac{1}{\tau_2}
\begin{pmatrix}
1 & \tau_1 \\ \tau_1 & |\tau|^2
\end{pmatrix} = e^\varphi
\begin{pmatrix}
1 & C_0 \\ C_0 & C_0^2 + e^{-2\varphi}
\end{pmatrix} \, .
\label{eq:EFTIIB}
\end{equation}
The dictionary for the gauge fields is
\begin{equation}
\begin{aligned}
	A_\mu{}^\alpha &= \hat{C}_{\mu s}{}^\alpha \,, \qquad
	B_{\mu\nu}{}^{\alpha, s} = \hat{C}_{\mu\nu}{}^{\alpha} 
		+ A_{[\mu}{}^s \hat{C}_{\nu]s}{}^\alpha \,,  \\
	C_{\mu\nu\rho}{}^{\alpha\beta, s} &= \epsilon^{\alpha\beta} \hat{C}_{\mu\nu\rho s} 
		+ 3 \hat{C}_{[\mu|s|}{}^{[\alpha} \hat{C}_{\nu\rho]}{}^{\beta]} 
		- 2 \hat{C}_{[\mu|s}{}^{\alpha} \hat{C}_{\nu|s|}{}^{\beta} A_{\rho]}{}^s \,,  \\
	D_{\mu\nu\rho\sigma}{}^{\alpha\beta, ss} &= \epsilon^{\alpha\beta} 
		\left( \hat{C}_{\mu\nu\rho\sigma} 
			+ 4 A_{[\mu}{}^s \hat{C}_{\nu\rho\sigma]s} \right) 
			+ 6 \hat{C}_{[\mu\nu}{}^{[\alpha} \hat{C}_{\rho|s|}{}^{\beta]} A_{\sigma]}{}^s \,.
\end{aligned}
\end{equation}

Above we have summarized the rules for showing the equivalence of the $\G$ EFT to both 11-dimensional supergravity and 10-dimensional type IIB supergravity. Let us now elaborate on the connection to F-theory, rather than just type IIB supergravity.

\begin{wrapfigure}{r}{0.5\textwidth}
  \begin{center}
    \includegraphics[width=0.48\textwidth]{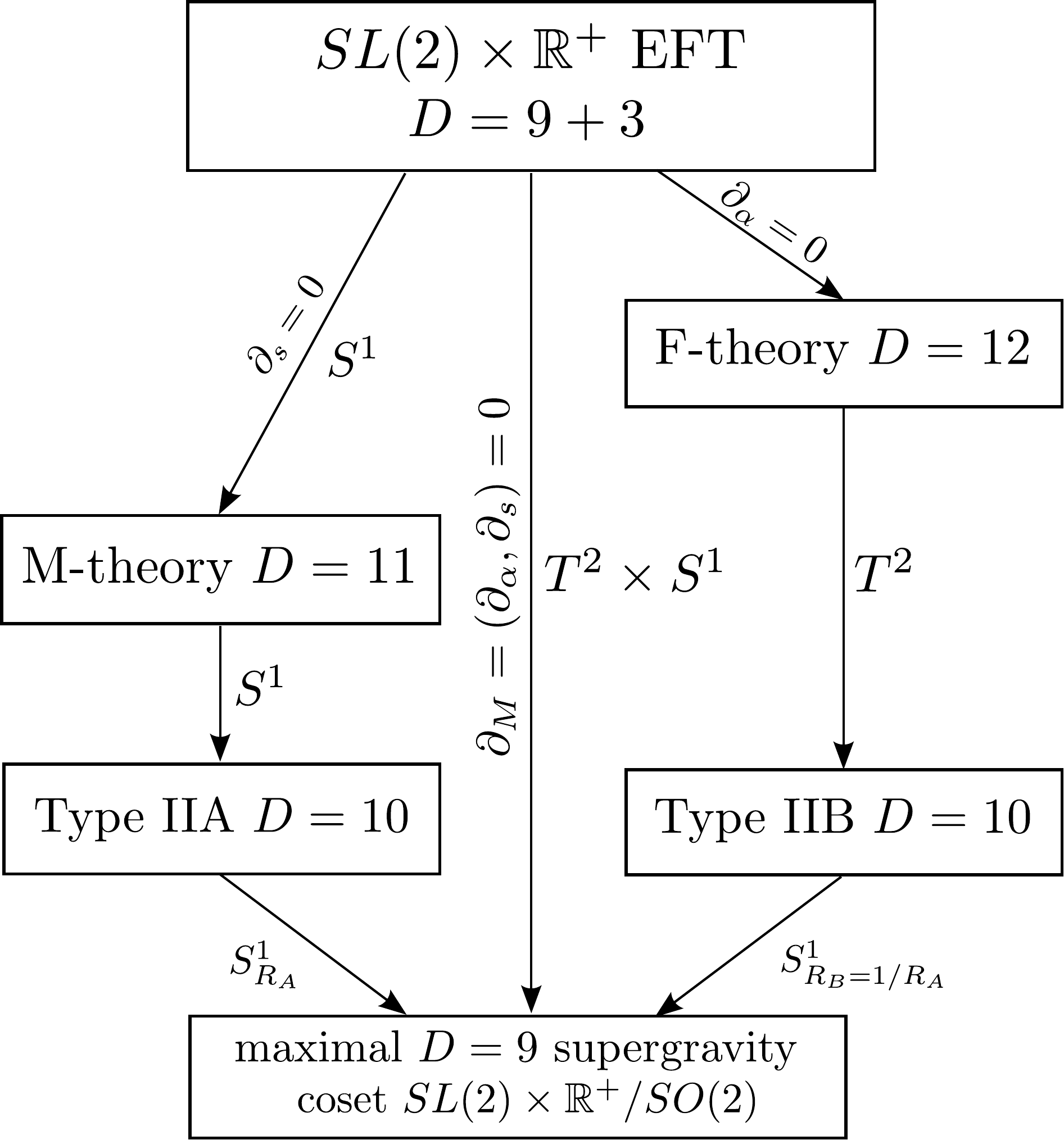}
  \end{center}
  \caption{The relation of supergravity theories in nine, ten, eleven and twelve dimensions. Note here $D$ denotes the overall dimensionality of the theory.}
\end{wrapfigure}

What is F-theory? Primarily we will take F-theory to be a 12-dimensional lift of IIB supergravity that provides a geometric perspective on the $\mathrm{SL}(2)$ duality symmetry. It provides a framework for describing (non-perturbative) IIB vacua with varying $\tau$, in particular it is natural to think of sevenbrane backgrounds as monodromies of $\tau$ under the action of $\mathrm{SL}(2)$. Equivalently, there is a process for deriving non-perturbative IIB vacua from M-theory compactifications to a dimension lower. Crucially, singularities of the 12-dimensional space are related to D7-branes. We take this duality with M-theory to be the second key property of F-theory.

We usually view the 12-dimensional space of F-theory as consisting of a torus fibration of 10-dimensional IIB. The group of large diffeomorphisms on the torus is then viewed as a geometric realisation of the $\mathrm{SL}(2)$ S-duality of IIB.

In the $\G$ EFT a similar picture arises. This is because we take the group of large \emph{generalised diffeomorphisms} acting on the extended space to give the $\G$ duality group. %See \cite{Hohm:2012gk,Park:2013mpa, Berman:2014jba,Hull:2014mxa,Naseer:2015tia,Rey:2015mba,Chaemjumrus:2015vap} for progress on understanding the geometry of these large generalised diffeomorphisms.
The EFT is subject to a single constraint equation, the section condition, with two inequivalent solutions. One solution of the constraint leads to M-theory or at least 11-dimensional supergravity, and one leads to F-theory. Thus $\G$ EFT is a single 12-dimensional theory containing both 11-dimensional supergravity and F-theory, allowing us to naturally realise the M-theory / F-theory duality.

If we choose the IIB section, we can interpret any solutions as being 12-dimensional but with at least two isometries in the 12-dimensional space. These two isometries lead to the 2-dimensional fibration which in F-theory consists of a torus.

Finally, the fact that the generalised diffeomorphisms, not ordinary diffeomorphisms, play the key role here also allows one to use the section condition to ``dimensionally reduce'' the 12-dimensional $\G$ to 10-dimensional IIB (as well as 11-dimensional supergravity) as explained above. This explicitly shows how F-theory, interpreted as the $\G$ EFT, can be a 12-dimensional theory, yet reduce to the correct 11-dimensional and type IIB supergravity fields. The relation between the various theories in nine, ten, eleven and twelve dimensions is depicted in Figure 1.

\section{Sevenbranes}

In F-theory, a vital role is played by backgrounds containing sevenbranes. In this section we discuss some features of how one may view sevenbranes and their singularities in the context of the $\G$ EFT.

Sevenbrane solutions of type IIB supergravity have non-trivial metric and scalar fields $\tau$. From the point of view of EFT, all of these degrees of freedom are contained within the metric $g_{\mu \nu}$ and the generalised metric $\gM_{MN}$. Thus we may specify entirely a sevenbrane background by giving these objects. Below we will use the notation
\begin{align}
\dd s^2_{(9)} &= g_{\mu\nu}\dd x^\mu \dd x^\nu \, , &
\dd s^2_{(3)} &= \MM_{\alpha\beta}\dd y^\alpha \dd y^\beta + \MM_{ss} ( \dd y^s)^2\, ,
\end{align}
to specify the solutions. It is not obvious that one should view the generalised metric as providing a notion of line element on the extended space, so in a sense this is primarily a convenient shorthand for expression the solutions.

We consider a sevenbrane which is extended along six of the ``external directions'', denoted $\vec{x}_6$, and along $y^s$ which appears in the extended space. The remaining coordinates are time and the directions transverse to the brane which we take to be the polar coordinates $(r,\theta)$. In this language, the harmonic function of the brane is $H \approx h \ln[r_0/r]$.\footnote{The cut-off $r_0$ is related to the codimension-2 nature of the solution, i.e. we expect it to be valid only up to some $r_0$ as the solution is not asymptotically flat.} The solution can be specified by
\begin{equation}
\begin{aligned}
\dd s^2_{(9)} &= -\dd t^2 + \dd \vec{x}_{(6)}^{\, 2} 
						+ H \left(\dd r^2 + r^2 \dd \theta^2 \right)  \\
\dd s^2_{(3)} &= H^{-1}\left[(\dd y^1)^2 + 2 h\theta \dd y^1\dd y^2 + K (\dd y^2)^2 \right]
						+ (\dd y^s)^2  \\
{\Aa_\mu}^M &= 0 \, , \qquad K = H^2 + h^2 \theta^2 \, .
\end{aligned}
\label{eq:sevenbrane}
\end{equation}
If one goes around this solution in the transverse space changing $\theta=0$ to $\theta=2\pi$, the $2\times 2$ block $\MM_{\alpha\beta}$ goes to
\begin{equation}
\MM \rightarrow \Omega^T\MM\Omega \, , \qquad 
\Omega = 
\begin{pmatrix}
1 & 2\pi h \\ 0 & 1 
\end{pmatrix} \, .
\end{equation}
where the monodromy $\Omega$ is an element of $\mathrm{SL}(2)$.

Reducing this solution to the IIB section gives the D7-brane. By using \eqref{eq:genmetricIIB} and \eqref{eq:EFTIIB} one can extract the torus metric $\cH_{\alpha\beta}$ and the scalar $\phi$ of the $10=9+1$ split. From $\cH_{\alpha\beta}$ one then obtains the axio-dilaton, i.e. $C_0$ and $e^\varphi$. The external metric is composed with $\phi$ to give the 10-dimensional solution
\begin{equation}
\begin{aligned}
\dd s^2_{(10)} &= -\dd t^2 + \dd \vec{x}_{(6)}^{\, 2} 
					+ H \left(\dd r^2 + r^2 \dd \theta^2 \right)	+ (\dd y^s)^2  \\
C_0 &= h\theta \, , \qquad e^{2\varphi} = H^{-2}			
\end{aligned}
\label{eq:D7}
\end{equation}
which is the D7-brane. Exchanging $y^1$ and $y^2$ and flipping the sign of the off-diagonal term (this is an $\mathrm{SL}(2)$ transformation of the $\MM_{\alpha\beta}$ block) leads to a solution which reduces to the S7-brane. On the M-theory section the solution \eqref{eq:sevenbrane} corresponds to a smeared KK-monopole which can be written as
\begin{equation}
\dd s^2_{(11)} = -\dd t^2 + \dd \vec{x}_{(6)}^{\, 2} 
					+ H \left(\dd r^2 + r^2 \dd \theta^2	+ (\dd y^1)^2\right)
					+ H^{-1}\left[\dd y^2 + h\theta\dd y^1\right]^2\, .
\label{eq:smearedKK}
\end{equation}
To see this more clearly, consider the usual KK-monopole in M-theory, which has three transverse and one isometric direction, the Hopf fibre. If this solution is smeared over one of the transverse directions to give another isometric direction one arrives at the above solution (where $(r,\theta)$ are transverse and $(y^1,y^2)$ are isometric). Therefore the M/IIB-duality between smeared monopole and sevenbrane relates the first Chern class of the Hopf fibration to the monodromy of the codimension-2 object. 

We have now seen how the $\mathrm{SL}(2)$ doublet of D7 and S7 is a smeared monopole with its two isometric direction along the $y^\alpha$ in the extended space. This can be generalized to give $pq$-sevenbranes in the IIB picture where the isometric directions of the smeared monopole correspond to the $p$- and $q$-cycles. The external metric is the same as above, the generalized metric now reads
\begin{equation}
\begin{aligned}
\dd s^2_{(3)} &= \frac{H^{-1}}{p^2+q^2}\bigg\{
					\left[p^2H^2 + (ph\theta - q)^2\right](\dd y^1)^2 
					+ \left[(p + qh\theta)^2 + q^2H^2\right](\dd y^2)^2  \\
	&\hspace{3cm} - 2\left[(p^2-q^2)h\theta + pq(K-1)\right] \dd y^1\dd y^2 \bigg\}
					+  (\dd y^s)^2  \, .
\end{aligned}
\label{eq:pqsevenbrane}
\end{equation}
The two extrema are $p=0$ which gives the D7 and $q=0$ which gives the S7. 

As for all codimension-2 objects, a single D7-brane should not be considered on its own. To get a finite energy density, a configuration of multiple sevenbranes needs to be considered. Introducing the complex coordinate $z=re^{i\theta}$ on the two-dimensional transverse space, such a multi-sevenbrane solution in EFT reads
\begin{equation}
\begin{aligned}
\dd s^2_{(9)} &= -\dd t^2 + \dd \vec{x}_{(6)}^{\, 2} + \tau_2|f|^2\dd z \dd \bz  \\
\dd s^2_{(3)} &= \frac{1}{\tau_2}\left[|\tau|^2 (\dd y^1)^2 
						+ 2\tau_1 \dd y^1\dd y^2 + (\dd y^2)^2 \right]	+(\dd y^s)^2 
\end{aligned}
\label{eq:multisevenbrane}
\end{equation}
where all the tensor fields still vanish. Instead of specifying a harmonic function on the transverse space, we now have the holomorphic functions $\tau(z)$ and $f(z)$. Their poles on the $z$-plane correspond to the location of the sevenbranes. One usually takes 
\be
\tau = j^{-1} \left( \frac{P(z)}{Q(z)} \right) \,,
\ee
where $P(z)$ and $Q(z)$ are polynomials in $z$ and $j(\tau)$ is the j-invariant. The roots of $Q(z)$ will give singularities, which in the IIB section give the locations of the sevenbranes. The configuration in this case consists of the metric 
\be
\dd s^2_{(10)} = -\dd t^2 + \dd \vec{x}_{(6)}^{\, 2} + \dd y_s^2 + \tau_2|f|^2\dd z \dd \bz  
\ee
together with the scalar fields encoded by $\tau$. Meanwhile in the M-theory section, one finds a purely metric background,
\be
\dd s^2_{(11)}  = - \dd t^2 + \dd \vec{x}_{(6)}^{\, 2}
+ \tau_2|f|^2\dd z \dd \bz 
 + \tau_2 ( \dd y^1 )^2 
+ \frac{1}{\tau_2} \left( \dd y^2 + \tau_1 \dd y^1 \right)^2 \,.
\ee
This retains the singularities at the roots of $Q$ at which $\tau_2 \rightarrow i \infty$. 

%A crucial point about singularities in F-theory is that they are seen as an origin for nonabelian gauge symmetries. These arise from branes wrapping vanishing cycles at the singularities. At this point we will not examine the details of this for our $\G$ EFT but instead we can point to interesting recent work in DFT. The authors of \cite{Aldazabal:2015yna} construct the full nonabelian gauge enhanced theory in DFT corresponding to the bosonic string at the self-dual point. One would hope that one could apply a simlar construction to produce a gauge enhanced EFT with non-abelian massless degrees of freedom coming from wrapped states on vanishing cycles.

In the original paper \cite{Berman:2015rcc} other solutions in the $\G$ EFT such as waves, monopoles, strings and fivebranes where also considered similar to the approach in \cite{Berman:2014jsa,Berman:2014hna,Berkeley:2014nza} for higher EFTs.

\section{Outlook}

We are aware that the approach of most practitioners in F-theory that has yielded so much success over a number of years has been through algebraic geometry. It is doubtful if the presence of this action can help in those areas where the algebraic geometry has been so powerful. We do hope though that it may provide some complementary techniques given that we now have a description in terms of 12-dimensional degrees of freedom equipped with an action to determine their dynamics. 

One question people have tried to answer is the theory on a D3-brane when $\tau$ varies. This might be computable in this formalism using a Goldstone mode type analysis similar to that in  \cite{Berkeley:2014nza} where this method was used to determine string and brane effective actions in DFT and EFT. 

A useful result from this formalism would be to show why elliptic Calabi-Yau are good solutions to the 12-dimensional theory. This would likely involve the construction of the supersymmetric version of the $\G$ EFT in order to study the generalised Killing spinor equation. Another interesting area of investigation would be the heterotic/type II duality, where we should then consider EFT on a K3 background and relate this to heterotic DFT \cite{Hohm:2011ex,Grana:2012rr}. This has been investigated recently in \cite{Malek:2016vsh}.

An interesting consequence of our work is that it shows how F-theory fits into a general picture of EFT with various $\Edd$ groups. One might be then be inspired to consider far more general backgrounds with higher dimensional fibres and with monodromies in $\Edd$ and so one would not just have sevenbranes but more exotic objects (of the type described in \cite{deBoer:2012ma}). 

%More recent work in this direction has appeared where one takes the fibre to be $K3$ and then one has a U-duality group act on the $K3$ \cite{Braun:2013yla,Candelas:2014jma,Candelas:2014kma} in a theory sometimes called G-theory. 

One can also consider more general types of reduction than the simple fibrations described here such as Scherk-Schwarz type reductions that yield gauged supergravities. Such a reduction on the F-theory torus -- this makes no sense from the IIB perspective but it does from the point of view of the $\G$ EFT -- leads to the massive Type IIA Romans' supergravity \cite{Cassani:2016ncu,Ciceri:2016dmd}.

A further quite radical notion would be the EFT version of a T-fold where we only have a local choice of section so that the space is not globally described by type IIB or M-theory. One could have a monodromy such that as one goes round a one-cycle in nine dimensions and then flips between the IIB section and M-theory section. This would exchange a wrapped membrane in the M-theory section with a momentum mode IIB section just as a T-fold swaps a wrapped string with a momentum mode. Note, this is not part of the $\G$ duality group and thus is not a U-fold. This is simply because with two isometries one has a $\mathbb{Z}_2$ choice of section that one can then twist.

\end{document}